\documentstyle[amssymb,aps,multicol,prl]{revtex}
\input{graphicx}

\draft

\begin{document}

\title{Vortex nanoliquid in high-temperature superconductors}

\author{S. S. Banerjee$^1$, S. Goldberg$^1$, A. Soibel$^2$, Y. Myasoedov$^1$, M.
Rappaport$^1$, E. Zeldov$^1$, F. de la Cruz$^3$, C. J. van der Beek$^4$, M.
Konczykowski$^4$, T. Tamegai$^5$, and V. M. Vinokur$^6$}

\address{$^1$Department of Condensed Matter Physics, Weizmann Institute of Science,
Rehovot 76100, Israel}
\address{$^2$Jet Propulsion Laboratory, California Institute of Technology, Pasadena,
California 91109}
\address{$^3$Instituto Balseiro and Centro At\'{o}mico Bariloche, CNEA, Bariloche, 8400, Argentina}
\address{$^4$Laboratoire des Solides Irradi\'{e}s, CNRS UMR 7642 and CEA-CMS-DRECAM,
Ecole Polytechnique, 91128 Palaiseau, France}
\address{$^5$Department of Applied Physics, The University of Tokyo, Hongo, Bunkyo-ku,
Tokyo 113-8656, Japan}
\address{$^6$Materials Science Division, Argonne National Laboratory, Argonne,
Illinois 60439}
\date{\today}
\maketitle

\begin{abstract}
Using a differential magneto-optical technique to visualize flow of transport
currents, we reveal a new delocalization line within the reversible vortex liquid
region in the presence of a low density of columnar defects. This line separates a
homogeneous vortex liquid, in which all the vortices are delocalized, from a
heterogeneous ``nanoliquid" phase, in which interconnected nanodroplets of vortex
liquid are caged in the pores of a solid skeleton formed by vortices pinned on
columnar defects. The nanoliquid phase displays high correlation along the columnar
defects but no transverse critical current.
\end{abstract}


\begin{multicols}{2}
Columnar defects (CDs) created in high-temperature superconductors by high-energy
heavy ion irradiation act as very efficient pinning centers for vortices aligned
along the CDs. The effects of this correlated disorder on the vortex properties
have been investigated extensively mainly in the limit of relatively high density
of CDs where the enhancement in the critical current for practical applications is
most significant and the vortex matter is known to form the Bose glass phase
\cite{Blatter,Nelson,Radzihovsky,Kees}. In contrast, our understanding of the
opposite limit, in which sparse CDs perturb a higher concentration of
vortices, is still incomplete. Recent experimental studies \cite{Banerjee,Menghini}
have shown that in this case the vortex matter can no longer be regarded as a
homogeneously pinned medium. Instead, an intrinsically heterogeneous structure with
two distinct sub-systems of vortices with very different characteristic energies is
formed: vortices residing on CDs are strongly pinned and create a rigid skeleton,
while the interstitial vortices form relatively ordered nanocrystals intercalated
within the voids of this porous skeleton. This experimental finding of a ``porous"
vortex solid has contributed to a series of recent theoretical studies and
numerical simulations \cite{Nonomura,Dasgupta,Tyagi,Lopatin,Rodriguez,Morozov}.
Moreover, it was found that the melting line $B_m(T)$ of the ``porous" vortex solid
has an unconventional behavior showing a pronounced kink \cite{Banerjee,Menghini}.
This has led to the suggestion of the possible existence of two different liquid
phases: at fields below the kink a usual homogeneous vortex liquid is formed upon
melting, whereas at fields above the kink the ``porous" solid melts into a
heterogeneous state in which liquid vortices coexist with a skeleton of pinned
vortices. It was proposed that in such a case a new transition line separating
these two liquid phases should be expected \cite{Banerjee,Radzihovsky,Lopatin}.
Using a new imaging technique we present here the first experimental evidence for
such a transition line within the reversible vortex liquid phase that separates a
heterogeneous ``nanoliquid" from a homogeneous liquid. One of the interesting
properties of this novel nanoliquid phase is the absence of in-plane
dissipationless superconductivity, while the coherence along the c-axis is
apparently still preserved.

\begin{figure}
\includegraphics[width=2.0in,height=3.0in,angle=-90]{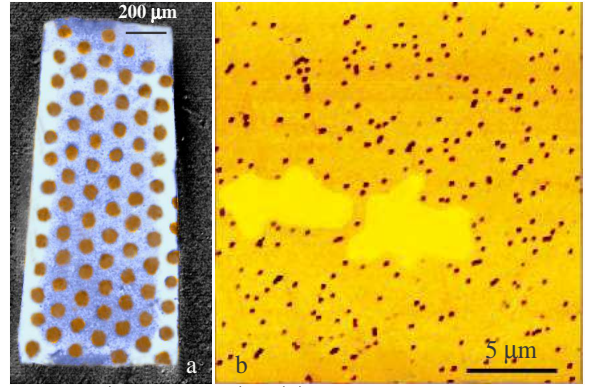}
\caption{(color online). (a) DMO image of melting in a BSCCO crystal at $B = 45$ G, $T = 79.75$ K,
and $T$ modulation of 0.3 K. All the pristine regions of the sample are in the
liquid phase (blue) while the circular regions irradiated with $B_{\phi} = 60$ G
are still solid (brown). (b) AFM $20 \times 20$ $\mu$m$^2$ image of an etched mica
irradiated along with BSCCO crystals ($B_{\phi} = 20$ G). The dark spots reveal the
locations of CDs while the bright yellow regions are examples of two voids in the
distribution of columns.}
\end{figure}

The CDs were created along the c-axis in  Bi$_2$Sr$_2$CaCu$_2$O$_8$ (BSCCO) single
crystals ($T_c \simeq 90$ K) using low irradiation doses of 1 GeV Pb ions at GANIL,
France. The samples were covered with 50 $\mu$m thick stainless steel mask with a
triangular array of 90 $\mu$m apertures. The mask blocks the ions and as a result
CDs are formed in the BSCCO crystals only within the 90 $\mu$m apertures. Figure 1a
shows the melting process in one of the BSCCO crystals of about $1.9 \times 0.8
\times 0.025$ mm$^3$ visualized by differential magneto-optical imaging (DMO) for
magnetic fields H $\parallel$ c-axis. The first-order melting transition (FOT) is
detected by observing a step in the equilibrium local magnetization
\cite{Pastoriza,Zeldov,Schilling}, which appears as bright paramagnetic regions in
the DMO images \cite{Banerjee,Soibel1}. For clarity, the liquid and solid regions
are colored in blue and brown respectively. In Fig. 1a all the pristine regions of
the sample have melted already (blue) while the 90 $\mu$m irradiated regions are
still solid (brown). Figure 2 shows the $B - T$ phase diagram with the
corresponding melting lines in the pristine, $B_m^0(T)$, and in the irradiated,
$B_m^{CD}(T)$, regions with $B_{\phi} = 60$ G ($B_{\phi} = n\phi_0$, n is the areal
density of CDs, and $\phi_0$ is the flux quantum). The use of the irradiation mask
allows accurate determination of the difference in the melting temperature in
adjacent irradiated and pristine regions in the same sample. Figure 2 shows that
the irradiated regions melt at substantially higher temperatures while strikingly
preserving the first-order nature of the transition \cite{Banerjee}. With
decreasing temperature along the $B_m^{CD}(T)$ line the strength of the FOT in the
irradiated regions diminishes, becoming weakly first-order and eventually turning
into a second order transition at a temperature that is irradiation dose and sample
dependent \cite{Banerjee,Khaykovich}. The open circles in Fig. 2 were obtained
using field modulation DMO while the solid circles using temperature modulation.

\begin{figure}
\includegraphics[width=2.5in,height=3.0in,angle=-90]{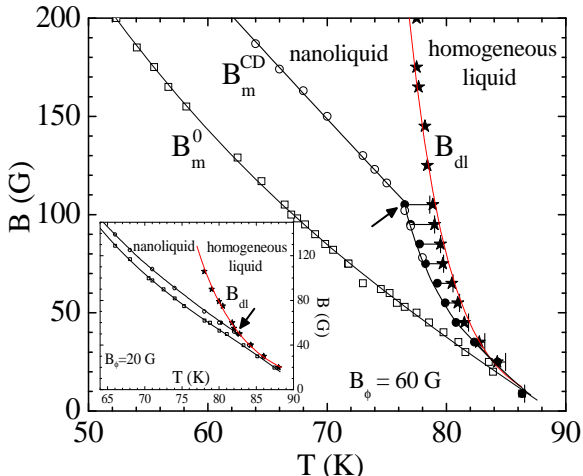}
\caption{(color online). $B-T$ phase diagram in BSCCO: $B_m^0(T)$ $-$ pristine melting,
$B_m^{CD}(T)$ $-$ melting line with CDs with $B_{\phi} = 60$ G, and $B_{dl}(T)$ $-$
delocalization line. The black lines are guides to the eye and the red $B_{dl}(T)$
line is a fit to exp$(-T/T_0)$. The arrow indicates the location of the kink in the
$B_m^{CD}(T)$ line. Inset: phase diagram of a $B_{\phi} = 20$ G irradiated sample.}
\end{figure}

We now focus our attention only on the irradiated regions which are macroscopic on
the scale of the intervortex distance. Figure 1b shows an AFM image of an etched
mica irradiated along with the crystals, which reveals the typical random
distribution of CDs following Poisson statistics. The CDs form a dense network or
matrix that contains numerous voids that are free of columns. Two such voids, or
pores, are outlined schematically by the light color in Fig. 1b. The CDs act as
strong pinning centers for the vortices and therefore at low $B$ all the vortices
reside on CDs. When the field approaches the matching field $B_{\phi}$ most of the
CDs become occupied by vortices. This network of pinned vortices forms a rigid
skeleton, while upon further increase of $B$, the extra vortices create ordered
nanocrystals within the pores of the skeleton, as revealed by magnetic decoration
studies \cite{Menghini}.

\begin{figure}
\includegraphics[width=2.5in,height=3.0in,angle=-90]{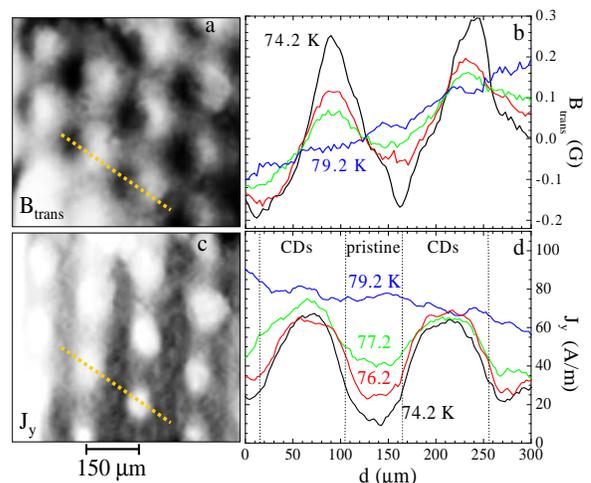}
\caption{(color online). (a) DMO image of magnetic field self-induced by a transport current of 50
mA in the central part of the sample at $B = 145$ G and $T = 74.2$ K. The $B_{\phi}
= 60$ G irradiated regions are in the nanoliquid phase. The bright and dark shades
correspond to positive and negative values of $B_{trans}$. (b) Profiles of
$B_{trans}$ along the yellow dashed line in (a) at different temperatures. (c)
Current distribution $J_y$ obtained by inversion of $B_{trans}$ image in (a). (d)
Profiles of $J_y$ along the yellow line in (c) at different $T$ with indicated
irradiated and pristine regions.}
\end{figure}

We now investigate the nature of the resulting liquid phase and search for a
possible new transition between two different liquid phases. Since our DMO
measurements do not show any FOT equilibrium magnetization steps within the liquid
region of the phase diagram, we have developed a new differential imaging technique
to search for a possible liquid-liquid transition. Instead of modulating the
applied field or temperature, a transport current with periodically alternating
polarity is applied to the sample and a corresponding differential image is
acquired and integrated, typically over a hundred cycles. As a result, the
self-induced magnetic field $B_{trans}$ generated by the transport current according
to the Biot-Savart law can be measured to a very high precision. With this
differential imaging we are able to determine the distribution of currents as low
as 0.1 mA, about three orders of magnitude improvement compared to previous methods
\cite{Indenbom}.

Figure 3a shows $B_{trans}$ in the central part of the sample resulting from 50 mA
transport current. This image displays the behavior at $T = 74.2$ K and $B = 145$
G, which is above the $B_m^{CD}(T)$ line. Here the entire sample is in the liquid
phase and no field variation between the irradiated and pristine regions is
observed by DMO using either field or temperature modulations. Surprisingly,
despite the fact that both the irradiated and pristine regions are liquid, strong
variations in $B_{trans}$ are found in Fig. 3b, revealing a highly
inhomogeneous distribution of transport current that reflects large differences in the vortex dynamics. Such an ability to resolve spatial variations of vortex dynamics within the liquid phase is a unique feature of this
new differential technique. Using the matrix inversion \cite{Wijngaarden,Brandt2} of
$B_{trans}$, a direct measure of the spatial distribution of the transport current
$J_y$ is obtained as shown in Figs. 3c and 3d. The current is strongly focused into
the irradiated regions reaching densities up to seven times higher than in the
unirradiated regions. This means that at this temperature the vortex liquid phase
in the irradiated regions is different from the pristine homogeneous liquid; it has
a much lower resistivity and the vortices experience a much higher effective
viscosity. Figures 3b and 3d also show the evolutions of $B_{trans}$ and $J_y$ with
increasing temperature. At $T_{dl}=79$ K the current distribution becomes uniform
quite abruptly and remains uniform at higher temperatures, showing that above
$T_{dl}$ the two liquids become identical. As discussed below, $T_{dl}$ is thus the
delocalization temperature separating a nanoliquid phase, in which interconnected
liquid droplets are caged by a solid skeleton, from a homogeneous liquid. Figure 2
shows the location of this delocalization line $B_{dl}(T)$ on the phase diagram,
which is our central result here.

We have also performed transport measurements simultaneously with the DMO imaging.
Figure 4 shows the measured four probe resistance $R$ of the sample as a function
of temperature at various fields. The arrows indicate the $T_{dl}(B)$ as derived
from the DMO self-field measurements. The resistance in the liquid phase of BSCCO
has Arrhenius behavior\cite{Palstra,Fuchs} as seen in the inset of Fig. 4. We
observe that in the homogenous liquid region, $R$ decreases smoothly with
decreasing temperature until $T_{dl}$ is reached. Below $T_{dl}$ a nanoliquid is
formed in the irradiated regions, significantly reducing $R$. In our special
geometry, however, the volume of the irradiated regions (brown regions in Fig. 1a)
is less than half of the total volume of the sample. Therefore we should expect
that, even if the resistance of the irradiated regions should become zero, the
total measured $R$ of the sample should drop approximately only by a factor of two,
since the pristine regions remain liquid (blue in Fig. 1a) and maintain highly
resistive Arrhenius behavior. Such a drop between two parallel Arrhenius lines is
indeed observed and is emphasized by the dotted lines in the inset of Fig. 4.
However, the striking observation is that the drop is by factor of more than $20$.
This behavior shows that the localization of the skeleton of vortices on the CDs
significantly enhances the c-axis correlation as follows: Because of the high
anisotropy of BSCCO the applied current flows in a very shallow layer underneath
the top surface of the crystal where the electrical contacts are
attached\cite{Busch,Safar2,Khaykovich2} (see Fig. 4 lower inset). The
characteristic depth of the current is inversely proportional to the effective
anisotropy $\gamma_{\rho} = (\rho_c / \rho_{ab})^{1/2}$, where $\rho_c$ and
$\rho_{ab}$ are the c-axis and in-plane resistivities. In BSCCO $\gamma_{\rho}
\approx 500$ and may change considerably depending on oxygen doping and
temperature. $\rho_c$ is determined by the degree of relative fluctuations and
misalignment between the vortex pancakes in the adjacent CuO$_2$ planes. The
formation of the rigid skeleton aligns the pancake stacks residing on the CDs and
also reduces the fluctuations of the vortices residing in the voids. As a result
the c-axis correlation and the Josephson coupling between the pancakes in the
nanoliquid phase should be significantly enhanced, giving rise to a large reduction
in $\rho_c$ and anisotropy as compared to the homogeneous liquid. Consequently, the
nanoliquid in the irradiated regions allows the current to penetrate much deeper
into the crystal, thereby significantly increasing the effective thickness of the
sample and reducing $R$. Since $R$ drops by a factor of more than twenty, instead
of a factor of two, we conclude that the nanoliquid is at least an order of
magnitude more isotropic than the homogenous liquid.

\begin{figure}
\includegraphics[width=2.5in,height=3.2in,angle=-90]{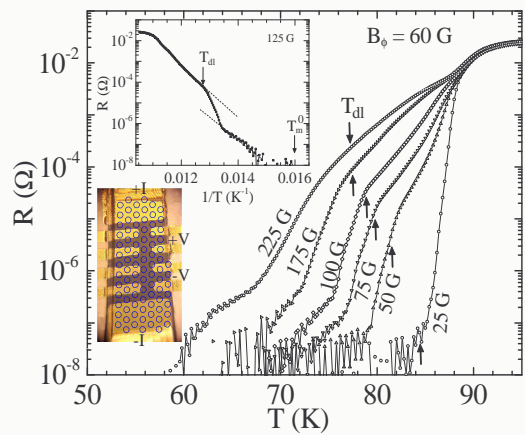}
\caption{(color online). Sample resistance vs. $T$ at various fields. The arrows indicate the
delocalization temperatures $T_{dl}$ as derived from DMO self-field data. Top
inset: example of Arrhenius plot of $R(T)$ at $B = 125$ G with indicated $T_{dl}$
and pristine melting $T_m^0$. Bottom inset: image of the gold contacts on the
crystal with irradiated regions shown schematically by the circles.}
\end{figure}

We now discuss the phase boundaries in Fig. 2 from the symmetry breaking
standpoint.  There are two symmetries that are broken upon freezing into the vortex
solid phase\cite{Blatter}: the formation of the Abrikosov lattice within the
nanocrystals locally breaks the continuous translational symmetry of the liquid,
and the appearance of the supercarrier density and the resulting ability to sustain
dissipationless currents along the vortices breaks the longitudinal gauge symmetry.
At $B_m^{CD}(T)$ the nanocrystals melt into nanodroplets, restoring the local
transverse translational symmetry. The rigid skeleton, however, remains intact due
to the strong pinning by the CDs, thus preserving the longitudinal c-axis
correlation along the CDs up to the $B_{dl}(T)$ line. The resulting nanoliquid
phase, consisting of interconnected nanodroplets caged in a porous skeleton, has a
unique property of extended c-axis coherence together with the absence of in-plane
critical current.

Several recent theoretical studies and numerical
simulations\cite{Nonomura,Lopatin,Rodriguez} indeed find evidence for
superconducting coherence along the c-axis in such an intermediate liquid phase. More
specifically, using the analogy between vortices and two-dimensional
bosons\cite{Blatter,Nelson}, an intermediate superfluid state in which the
superfluid condensate of bosons (vortex liquid) coexists with localized bosons
(vortices pinned on CDs) has been predicted\cite{Lopatin}. With increasing $T$ the
intermediate vortex liquid system (nanoliquid) undergoes a sharp delocalization at
$B_{dl}(T)$ above which all the vortices become delocalized (homogeneous liquid),
thereby rendering the system incapable of sustaining any supercurrents along the
vortices. This results in $B_{dl}(T) \propto$ exp$(-T/T_0)$, where
$T_0$ describes the effective depinning energy scale for vortex delocalization
\cite{Lopatin}, estimated to be $T_0 \sim 5$ K for typical parameters of BSCCO. The red
lines in Fig. 2 are fits to the data resulting in $T_0 \approx 7.7$ K for the
$B_{\phi} = 60$ G sample and $T_0 \approx 13.9$ K for the $B_{\phi} = 20$ G crystal
(inset) which had a slightly higher $T_c$. We have also studied a crystal with
$B_{\phi} = 47$ G (not shown) giving $T_0 \approx 7.3$ K. In all three cases the
theoretical temperature dependence provides a very good description of the
experimental $B_{dl}(T)$ line. Note, however, that the theory does not take into
account the discreteness associated with the pancake vortices in BSCCO. An alternative
estimate, based on the Berezinskii-Kosterlitz-Thouless like scenario of delocalization
of two-dimensional pancake vortices \cite{Kees}, also yields an exponential
dependence of $B_{dl}(T)$ with proper parameters. Thus, we surmise that the
localization of vortices on the porous skeleton depresses the c-axis resistivity
and breaks the longitudinal gauge symmetry in the nanoliquid, which is restored
only above the $B_{dl}(T)$ line. It is commonly believed that the formation of the
solid vortex lattice causes simultaneous breaking of the two symmetries across the
FOT. Our findings here show the first phase diagram identifying two separate phase
boundaries $B_m^{CD}(T)$ and $B_{dl}(T)$ across which the two symmetries of the
vortex matter are apparently broken separately. This unique property is the result
of the heterogeneous nature of the vortex nanoliquid.

It is interesting to note that at fields below the kink (marked by arrows in Fig.
2) the two symmetries are apparently broken simultaneously, where the $B_m^{CD}(T)$
line closely follows the delocalization line $B_{dl}(T)$. At these fields the
nanocrystals contain just a few vortices which remain solid as long as the caging
skeleton exists. The delocalization of the skeleton at $B_{dl}(T)$ thus causes
instantaneous melting of the entire lattice. This is a novel kind of vortex lattice
melting in which delocalization \cite{Kees,Lopatin} of one sub-system of vortices
induces a FOT in the entire heterogeneous system. Note that the finite distribution
of the nanocrystal sizes results in broadening of the FOT, as indicated by the
error bars in Fig. 2. The FOT apparently commences upon melting of the regions with
larger crystallites while the global homogeneous liquid is established only upon
melting of the smallest of the crystallites. Interestingly, this broadening is
hardly resolvable for lower irradiation doses, as seen in the inset of Fig. 2.

In conclusion, a low density of CDs is shown to significantly perturb the vortex
liquid resulting in formation of an intermediate nanoliquid phase in which liquid
vortex droplets are caged by vortices localized on CDs. We find evidence for a new
delocalization line at which the nanoliquid transforms into a homogeneous vortex
liquid forming a ``Y" shaped phase diagram. The nanoliquid phase has unique
properties resembling the long sought disentangled vortex liquid \cite{Feigelman}
with high c-axis correlation but zero transverse critical current. We conjecture
that the nanoliquid phase with coexisting localized and delocalized particles may
be generic to strongly interacting systems in the presence of disorder.

We thank M. Menghini, Y. Fasano, E. H. Brandt, and V. Geshkenbein for stimulating
discussions and R. J. Wijngaarden for providing the matrix inversion algorithm.
This work was supported by the Israel Science Foundation Center of Excellence, by
the German-Israeli Foundation G.I.F., and by the Minerva Foundation, Germany. The
work of FdlC was partially supported by Fundacion Antorchas and by ANPCYT PICT99-5117, and of VV by the U.S. Department of Energy, Office of Science. TT
acknowledges the support of the Grant-in-Aid for Scientific Research from the
Ministry of Education, Culture, Sports, Science, and Technology, Japan. EZ
acknowledges the support by the US-Israel Binational Science Foundation (BSF) and
by the Wolfson Foundation.

\end{multicols}

\end{document}